\PassOptionsToPackage{svgnames}{xcolor}
\documentclass[preprint,journal]{vgtc}       % preprint (journal style)
\ieeedoi{10.1109/TVCG.2023.3326523}

\onlineid{1655}

\vgtccategory{Research}

\vgtcpapertype{evaluation}

\title{Perception of Line Attributes for Visualization}

\author{%
\authororcid{Anna Sterzik}{0000-0002-0544-5397},
\authororcid{Nils Lichtenberg}{0000-0002-1008-803X},
\authororcid{Jana Wilms}{0009-0003-4078-5287},
\authororcid{Michael Krone}{0000-0002-1445-7568},
\authororcid{Douglas W. Cunningham}{0000-0003-1419-2552},
and \authororcid{Kai Lawonn}{0000-0002-1511-4022} 
}

\authorfooter{
    \item Anna Sterzik, Jana Wilms, and Kai Lawonn are with University of Jena.\\
    E-mail: \{anna.sterzik, jana.wilms, kai.lawonn\}@uni-jena.de
    \item Nils Lichtenberg and Michael Krone are with University of Tübingen. \\
    E-mail: \{nils.lichtenberg, michael.krone\}@uni-tuebingen.de
    \item Douglas W. Cunningham is with Brandenburg University of Technology.
    E-mail: douglas.cunningham@b-tu.de
}

\abstract{%
Line attributes such as width and dashing are commonly used to encode information.
\review{
However, many questions on the perception of line attributes remain, such as how many levels of attribute variation can be distinguished or which line attributes are the preferred choices for which tasks.
}
We conducted three studies to develop guidelines for using stylized lines to encode scalar data.
\review{
In our first study, participants drew stylized lines to encode uncertainty information.
}
Uncertainty is usually visualized alongside other data.
Therefore, alternative visual channels are important for the visualization of uncertainty.
Additionally, uncertainty---e.g.,\ in weather forecasts---is a familiar topic to most people.
Thus, we picked it for our visualization scenarios in study 1.
\review{
We used the results of our study to determine the most common line attributes for drawing uncertainty: Dashing, luminance, wave amplitude, and width.
While those line attributes were especially common for drawing uncertainty, they are also commonly used in other areas.
}
In studies 2 and 3, we investigated the discriminability of the line attributes determined in study 1.
Studies 2 and 3 did not require specific application areas; thus, their results apply to visualizing any scalar data in line attributes.
\review{
We evaluated the just-noticeable differences (JND) and derived recommendations for perceptually distinct line levels.
}
We found that participants could discriminate considerably more levels for the line attribute width than for wave amplitude, dashing, or luminance.
}

\keywords{Line Drawings, Line Stylization, Perceptual Evaluation, Uncertainty Visualization.}

\teaser{
 \includegraphics[width=\linewidth]{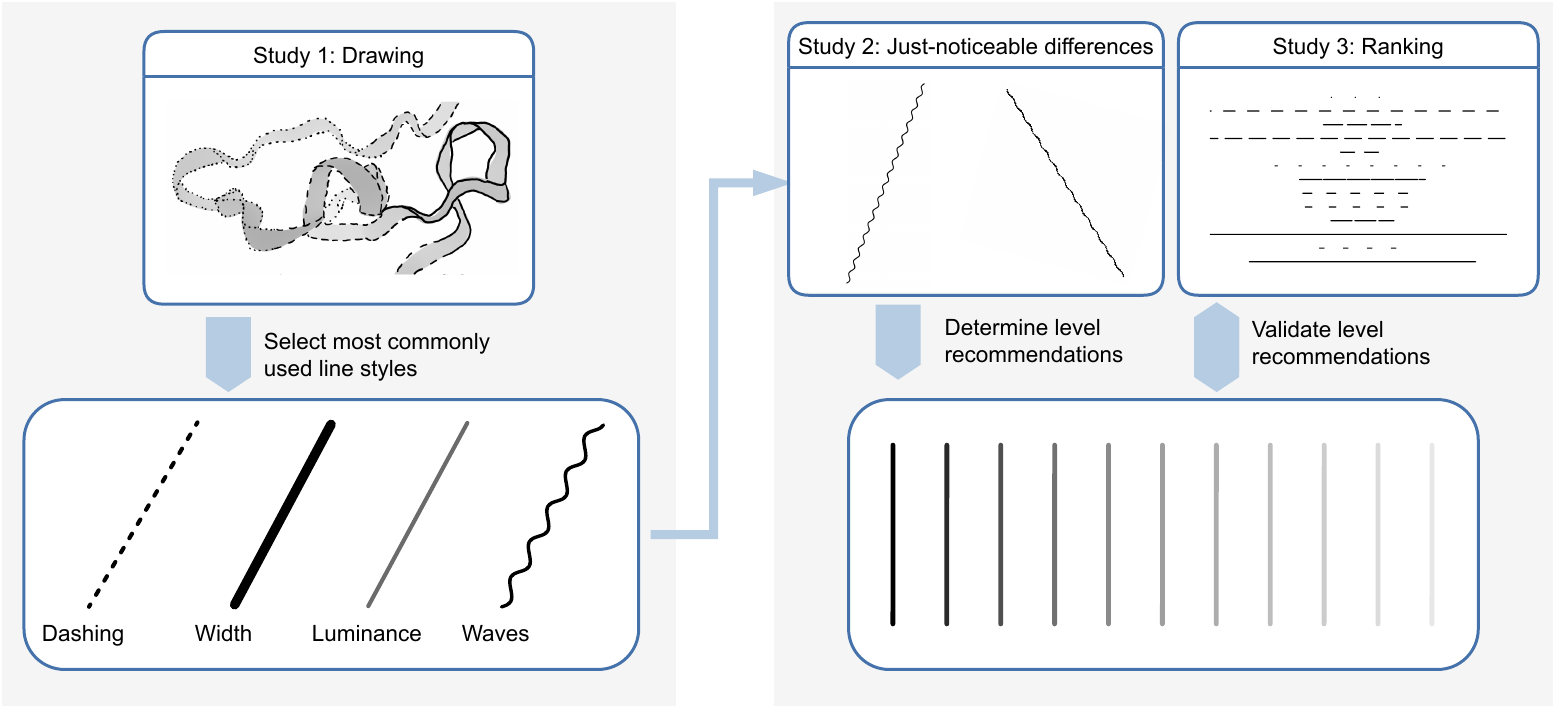}
 \centering
  \caption{
Workflow for investigating stylized lines for visualizing scalar data.
\review{
In the first study, people drew stylized lines to depict uncertainty data.
}
Based on the drawings, we selected the most commonly used line attributes.
For these line attributes, we determined the just-noticeable differences (JND) in study 2.
The JND were then used to derive level recommendations.
Finally, in study 3, we validated the level recommendations
using a ranking task.
}
\label{fig:teaser}
}

\graphicspath{{figs/}{figures/}{gfx/}{pictures/}{images/}{./}} % where to search for the images

\usepackage{todonotes}
\usepackage{svg}
\usepackage{subcaption}
\usepackage{float}
\usepackage{booktabs} % for nice tables
\usepackage{multirow} 
\usepackage{nicematrix} % for colored rows in table
\usepackage{gensymb}
\usepackage{import}%for pdf_tex
\usepackage{moresize}
\newcommand{\review}[1]{\textcolor{Black}{#1}}
\newcommand{\reviewtwo}[1]{\textcolor{Black}{#1}}
\usepackage{mathptmx}                  % use matching math font

\begin{document}

%%%%%%%%%%%%%%%%%%%%%%%%%%%%%%%%%%%%%%%%%%%%%%%%%%%%%%%%%%%%%%%%
%%%%%%%%%%%%%%%%%%%%%% START OF THE PAPER %%%%%%%%%%%%%%%%%%%%%%
%%%%%%%%%%%%%%%%%%%%%%%%%%%%%%%%%%%%%%%%%%%%%%%%%%%%%%%%%%%%%%%%

\firstsection{Introduction}

\maketitle

Lines as graphical primitives are omnipresent.
We can find them in maps, artistic line drawings, or visualizations.
\review{
To facilitate informed visualization design, we conducted three studies to evaluate the preferences and discriminability of stylized lines.
Stylized lines are intentionally modified lines where attributes such as width, color, or shape are modified to create a specific, often artistic, appearance.
Stylized lines can encode information by adjusting a specific attribute, such as luminance, according to a usually quantitative value.
For example, dark lines could encode certainty while lighter lines encode uncertainty.
Here, we focus on stylized lines for information encoding.}

First, we attempt to find popular line attributes by letting people draw lines themselves.
We developed a study with eight scenarios in which the participants were asked to create line drawings.
While stylized lines can display many types of data, the scenarios in our study needed to be easily understandable for the general public.
Additionally, the application area needs to be well-suited for encoding data in stylized lines.
Therefore, we chose uncertainty visualization as the use case for study 1.
Uncertain data should be familiar to the general public, as most people encounter uncertain data daily, e.g.,\ in weather forecasts or bus timetables.
Additionally, uncertainty is mostly displayed alongside other attributes.
Common visual channels, such as color, are often already occupied by these other attributes, and it is crucial to explore additional visual channels, such as line attributes.
Varying line attributes as indicators of uncertainty are popular~\cite{Guo2015, Gortler2018, Sterzik2022}.
A previous study by Boukhelifa et al.~\cite{Boukhelifa2012} compared the intuitiveness of the line attributes sketchiness and blur for the visualization of uncertainty.
\review{Sketchiness refers to perturbations of a line that mimic hand-drawn lines, i.e.,\ the perturbations are more irregular than waves.
The higher the deviations, the higher the sketchiness.}
They found sketchiness to be as intuitive as blur.
\review{
With our drawing study, we followed a more open approach.
Our evaluation was not limited to a set of pre-chosen line attributes.
The participants were unbiased; thus, we could evaluate popular line attributes and possibly new styles.
}

Using the results of study 1, we determined the four most commonly used line attributes for drawing uncertainty: luminance, dashing, waves, and width.
While using blur or fuzziness is often recommended as intuitive for \review{visualizing} uncertainty~\cite {MacEachren1992, Boukhelifa2012, MacEachren2012}, it was only used twice by our participants.
\review{
We believe that for most people, blur is harder to draw than, e.g.,\ dashing.
This might be a reason for the low prevalence of blurred lines.
They might still be intuitively associated with uncertainty.}

\review{%
For the line attributes most frequently used by our participants, we give guidance on discriminability.
}
While we chose the line attributes according to our study with an uncertainty visualization context,
neither of the other studies concerns uncertainty visualization in particular.
The results apply to encoding all types of scalar data in line attributes.
There are guidelines advising against the use of too many levels, for example, for width~\cite{Munzner2014}, but how many are too many?
\review{
We conducted a perceptual study to calculate the just-noticeable differences (JND) for luminance, wave amplitude, width, and two types of dashing.
The JND describes the amount of change to a stimulus along a given dimension that is needed, such that people can perceive that the stimulus changed~\cite{Cunningham2012}. 
}
Thus, based on the JND, we could determine a range of distinguishable levels for the line attributes.
\review{
To verify our chosen levels, we conducted a ranking study.
}
The participants ordered the levels from lowest to highest in drag-and-drop style tasks.
We confirmed that our level recommendations are distinguishable.
Therefore, they can be used as a starting point for designing visualizations based on line primitives.

To summarize, we developed guidelines for using stylized lines to encode scalar data.
\review{
We investigated how people draw stylized lines that represent uncertainty.
Luminance, dashing, waves, and width were the preferred choices for this purpose.
Furthermore, for these line attributes, we evaluated the JND and developed level recommendations, which were verified by a ranking study.}
The results for the level recommendations are relevant to all kinds of scalar data.
\section{Related Work}
In the following section, we will describe related work in the fields of line drawings, uncertainty visualization, and perceptual evaluations.

\subsection{Line Drawings}
\label{sec:line_drawings}
We mainly consider line drawings that represent surfaces.
However, other application areas can be found in information visualization.
In books that do not support colored graphics, for example, other line attributes need to be used to ensure discriminability.
Line drawing techniques applied to surfaces can be divided into \emph{silhouettes and contours}, \emph{feature lines}, and \emph{hatching}~\cite{Lawonn2018a}.
The techniques of the first and second category are usually used for stylizable lines.
Pioneering work was done by Appel~\cite{Appel1967}, Saito and Takahashi~\cite{Saito90}, and Dooley and Cohen~\cite{Dooley90}, who determined the silhouettes and contours and various line attributes for improved spatial impression. 
Feature lines are placed at spatially important regions to convey the shape of the surface.
The challenge is to define these important regions such that the lines represent the object well.
One of the most popular feature line techniques are \emph{suggestive contours} that extend traditional contours by considering the influence of nearby viewpoints.
% % %%
For more information on feature lines, we recommend the survey by Lawonn et al.~\cite{Lawonn2015Feature}.
\review{In hatched images, shading is approximated by closely spaced parallel lines.
An early hatching approach was presented by Deussen et al.~\cite{Deussen99},
more recent approaches using the GPU to improve performance were presented by Lawonn et al.~\cite{Lawonn_2013_CGF, Lawonn_2014_CGF}. 
While there are more examples of stylized lines for silhouettes, contours, and feature lines, different hatching styles have been used to encode information in medical scenarios~\cite{ Lawonn_2015_MICCAI, Lawonn2017}, where different hatch types encode the distance to a tumor.
A drawback of these methods was that they could not be applied to time-varying surfaces, i.e.,\ when the surface or distances to a focal object changed.
Therefore, Lichtenberg et al.~\cite{LICHTENBERG2018} proposed a real-time approach that can compute the hatch directly and also vary the density of the hatch.
They later improved their method to display different hatch strokes and contour styles that can be animated~\cite{Lichtenberg_2019_CGF}.}

Stylizing lines is another area of research.
Winkenbach and Salesin~\cite{Winkenbach1994} suggested imitating pen-and-ink illustrations with computer-generated lines.
They used stroke textures to achieve specific tones and textures in their line drawings.
A general challenge for line stylization methods is the generation of temporal-coherent stylized lines.
Recomputing the lines at every time frame leads to lines sliding or popping up~\cite{Benard2011}.
By propagating the lines from frame to frame, Kalnins et al.~\cite{Kalnins2003} provided a solution that works at interactive rates for models of medium complexity.
In non-interactive scenarios, the method by Buchholz et al.~\cite{Buchholz2011} generates a space-time surface that allows temporally coherent parameterizations.
Another approach for interactive scenarios was presented by Bénard et al.~\cite{Benard2012}.
Their active contours track and vectorize feature lines to generate stylizable brush paths for models of medium complexity.
For a more detailed overview of stylizing lines, we recommend the tutorial by Bénard et al.~\cite{Benard2019}.
\review{
The methods described above stylize lines for illustration purposes.
However, stylized lines can also be used for encoding information in the intensity of a varied line attribute.
Examples of line attributes used for information encoding include color or luminance, opacity, blur, width, wavelength or amplitude, and sketchiness~\cite{Guo2015, Gortler2018, Sterzik2022, Sterzik2023}.
In the following section about uncertainty visualization, some of the methods that use line attributes for uncertainty encoding are discussed.
}

\subsection{Uncertainty Visualization}
Uncertainty in Visualization ranges from erroneous input data to misinterpretation of the visualization by the viewer.
To cope with this situation, the domain of Uncertainty Visualization (UV) tries to quantify and visualize the uncertainty of data. 
Important goals of this discipline are to reduce impaired decision-making and to increase a viewer's trust in a visualization~\cite{sacha_role_2016}.
Geospatial Information Systems (GIS) probably have the longest history of UV in combination with spatial data~\cite{Pang2001}.
Employing ensemble data is a common approach to show that simulations representing a certain experiment can result in different final states~\cite{wang_visualization_2019}.
Therefore, ensemble data easily uncover that insight taken from simulations needs to be interpreted with simulation errors or uncertainties in mind.
Recent surveys confirm the importance of UV:
Kamal et al.~\cite{kamal_recent_2021} provide a general overview, while Gillmann et al.~\cite{gillmann_uncertainty-aware_2021} write about UV in medical visualization.

When adding uncertainty information to a visualization, that information needs to be communicated via a dedicated visual channel.
Especially if other information channels, like color, are exhausted, lines have the potential to open that additional channel.
Lechner~\cite{Lechner2020} investigates the usage of connecting lines, e.g.,\ in diagrams, to convey uncertainty while Gortler et al.~\cite{Gortler2018} use differently styled lines or contours to augment treemaps.
Stylized lines were also used by Strothotte et al.~\cite{Strothotte1999} to depict drawings of ancient architecture.
Other line-based effects have been used by Smit et al.~\cite{Smit2017} to communicate errors in the surfaces displayed in a surgical planning tool.
Zhao et al.~\cite{Zhao2019} provide a visualization system for public transport system uncertainties. 
In molecular visualization, line drawings to encode thermal vibration have been evaluated by Sterzik et al.~\cite{Sterzik2022, Sterzik2023}.

As shown by the references, line drawings were used for uncertainty visualization before. 
However, the feasibility of explicit line attributes to convey uncertainty has not been considered in an isolated environment but rather in the context of the respective application.

\subsection{Perceptual Research} 
While low-level psychophysical research has investigated the perception of lines for decades \cite{Granit1921}, most results are not directly applicable to more high-level tasks.
We aim to recommend levels for line attributes in visualizations.
Most visualization tasks are many perceptual steps away from pure low-level results.
Thus, we focus on high-level perceptual research.
Previous work by Cole et al.~\cite{Cole2008a} found that artists tend to use lines similar to computer-generated lines.
In their study, artists drew specific 3D shapes that were analyzed regarding their similarity to line drawing algorithms and other artists' drawings.
In further work, Cole et al.~\cite{Cole2009} found that line drawings are very effective in conveying shape.
MacEachren et al.~\cite{MacEachren2012} evaluated the intuitiveness and relative performance of several visual variables for uncertainty visualization.
In contrast to our work, they considered point symbol sets instead of lines during their evaluations.
Concerning intuitiveness, the participants rated a pre-selected set of visual variables regarding its intuitiveness.
In the second part of their study, they assessed how well participants can aggregate uncertainty information presented in a map-like fashion.
Boukhelifa et al.~\cite{Boukhelifa2012} investigated sketchiness as a visual variable for depicting uncertainty.
They compared sketchy lines with blurred, dashed, and grayscale (luminance) lines.
To determine the intuitiveness, they presented sketchy lines to the participants and asked them about their interpretations of the sketchiness.
In an additional experiment, they provided a list of possible interpretations of sketchiness from which the participants had to pick. 
The results of the second experiment were compared to similar results for blur.
In the second part of their study, they compared how accurate the perception of the line attributes is.
With the results of this evaluation, they tried to determine the number of discriminable levels for each line attribute.
The main difference to our evaluations is that our participants were unbiased towards the type of line attribute when drawing their uncertain line drawings and that we explicitly designed a study for determining discriminability.
Thus, we could determine more fine-grained distinct levels for the line attributes.
Sterzik et al.~\cite{SterzikTextures} constructed perceptually uniform levels for illustrative textures.
Illustrative textures, like stylized lines, can be used to convey information through visualizations.
In contrast to our JND-based approach, they used multi-dimensional scaling to construct perceptually uniform levels.
This approach has the advantage of characterizing the whole perceptual space but is unable to determine just noticeable differences.
Tak et al.~\cite{Tak2014} found that non-experts seem to perceive uncertainty as a distribution closely resembling the normal distribution.
The stimuli of their study use bands created with line primitives to display uncertainty.
The separability of two line attributes was studied by Guo et al.~\cite{Guo2015}.
They evaluated paired visual variables for representing uncertainty and another main attribute of graph edges.
Fuzziness, grain, and transparency seem to be robust choices for secondary attributes such as uncertainty.
Zahan et al.~\cite{Zahan2021} used stylized contour plots to visualize multivariate geospatial data.
They evaluated their custom designs in three crowd-sourced studies.
The designs used incorporated the line attributes width, hue, and luminance.
Sterzik et al. conducted two perceptual evaluations of line attributes for uncertainty visualization on molecular surfaces~\cite{Sterzik2022, Sterzik2023}.
They investigated how accurately people can distinguish between two different amounts of blur, dashing, luminance, sketchiness, and width.
They concluded that width and luminance robustly provide good results while dashing is very sensitive to the connectivity of the input lines.

\section{Overview of the Perceptual Studies}
Figure \ref{fig:teaser} displays the progression of our three research studies.
\review{
The first study investigates people's preferences for using specific line attributes for encoding uncertainty; the second and third study investigate the discriminability of several line attribute levels.
More specifically, study 1 aimed to understand how people draw lines and identify preferred line attributes for representing uncertainty through participant drawings in various scenarios.
In study 2, we conducted a Just Noticeable Difference (JND) evaluation to determine distinct levels based on the predetermined line attributes identified in study 1.
In study 3, we validated the estimated line levels through an online study, where participants sorted the levels for each line attribute, ensuring the practicality of the recommendations across diverse conditions.
Combining the results of studies 2 and 3, we attempt to find the maximum of discriminable line levels per attribute.
}
We received ethical approval for each study in this paper.
Studies 1 and 3 were without reimbursement.
For study 2, because of the long participation times, each participant was compensated with 15€.
\review{
The participants were mainly recruited using flyers and university-based mailing lists.
All three studies are independent of each other, so some participants participated in multiple studies, others in only one.
This does not introduce confounding effects, as study 1 concluded before the other two studies began, eliminating bias.
Participation in either of the other studies was neither beneficial nor detrimental to participation in studies 2 and 3.
}

\begin{figure}[tb]
\begin{subfigure}[t]{\linewidth}
    \centering
    \includegraphics[width=0.65\linewidth]{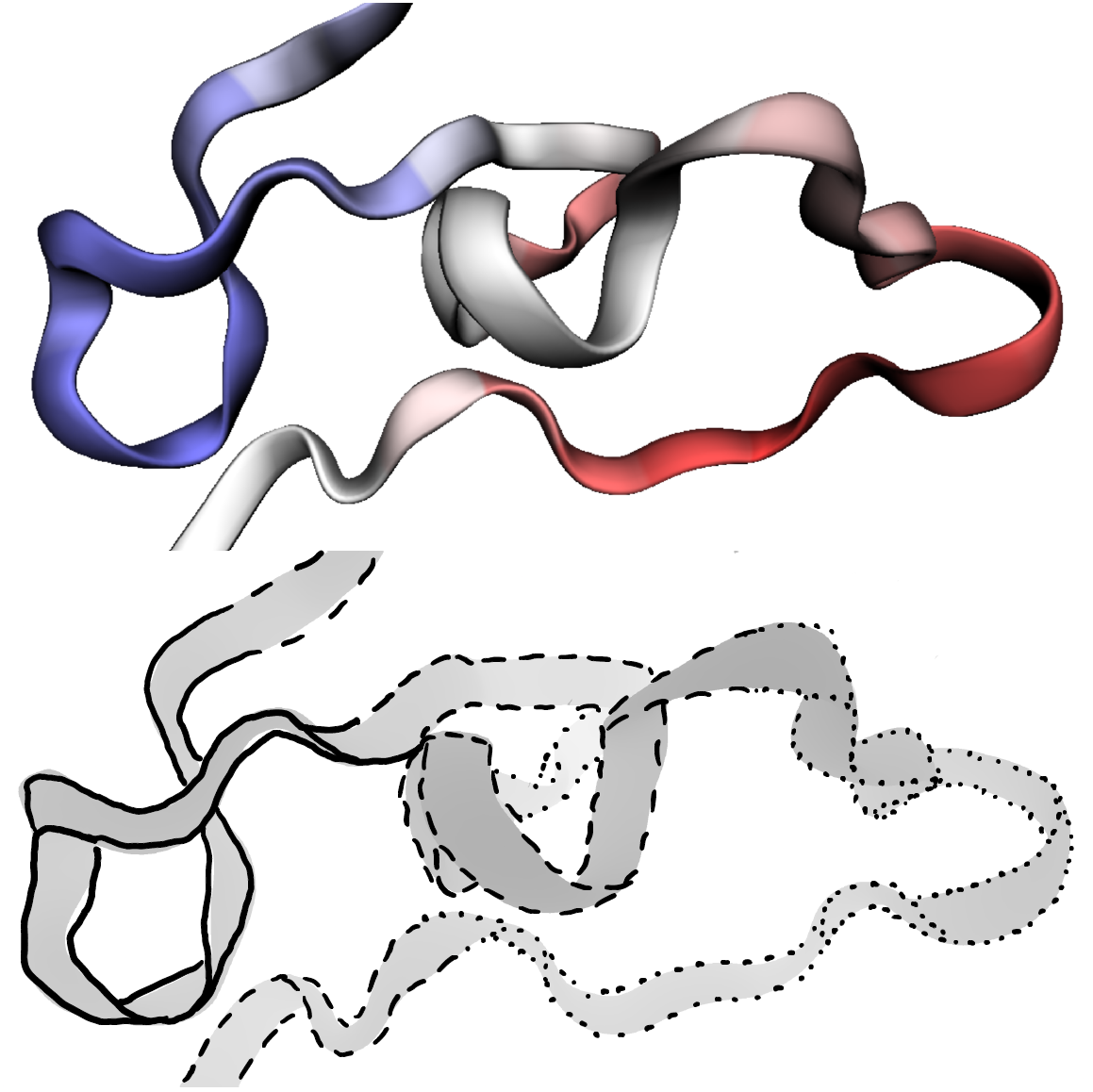}
    
    \caption{Molecular task.
    Two visualizations of a protein (Flagellin, PDB ID: 1IO1~\cite{1io1PDB, 1IO1}) created using VMD (Visual Molecular Dynamics)~\cite{VMD}) were given as input.
        In the upper visualization, the uncertainty is color-coded.
        The color map is typical for encoding flexibility in biochemistry, where blue means very stable and red means very flexible.
        In the lower visualization, the depth is encoded in the gray value, and no uncertainty information is given.
    The participants had to transfer the uncertainty from the upper image to the lower one by drawing lines.
    The image has been scaled down.
    \label{fig:mol-task}
    }
\end{subfigure}

\begin{subfigure}[t]{\linewidth}
    \centering
    \includegraphics[width= 0.8\linewidth]{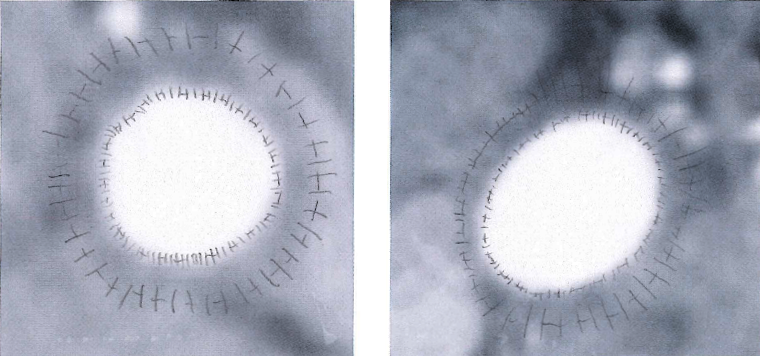}
    \caption{Blood vessel task.
    The inner and outer vessel walls had to be drawn onto the image.
    This is an example of one of the more uncommon line attributes found in our drawing study.
    Lines perpendicular to the suspected contour line of the vessel walls are used to indicate the extent of the uncertain area.
    The images have been scaled down.}
    \label{fig:error_bars}
\end{subfigure}
\caption{Example submissions for two tasks of the drawing study.}
\label{fig:example-submissions}
\end{figure}

\section{Drawing Stylized Lines}
\label{sec: drawing}

First, we investigated how people would \emph{draw} stylized lines.
\review{
By letting the participants draw the lines themselves, we determined popular line attributes for drawing uncertainty without limiting us to a predetermined selection.
Additionally, this allowed us to potentially find interesting line attributes that have not been considered by the visualization community yet.
}
We developed eight tasks in which the participants had to add lines in scenarios characterized by uncertain information.
The drawings could either be done with pencils on paper or with tablets.
\review{
The main purpose of this study was the identification of popular line attributes.
As such, we believe that the medium on which the lines were drawn is not crucial to answering this question.
In the submitted line drawings, we did not find any obvious differences between the drawings on paper and the digital submissions.
However, a larger sample size would be necessary to investigate this more closely.
}
The participants were asked to refrain from using color to elicit more varied responses and leave the color channel open for other information.

\subsection{Experimental Setup}
\review{
The first three participants participated in a pilot version of the study.
The feedback from the pilot study lead to very
minor modifications---e.g.,\ a lower saturation in one of the images---to the main study.
As such, we combined the answers
from the pilot study and the main study. 
Participation was possible in-person and online.
Participants who used paper were instructed to only use pencils for their line drawings, while online participants were instructed to refrain from using colors and could use a tablet.
Twelve participants submitted pencil-drawn lines, five submitted digital lines, and one participant submitted ballpoint pen drawings, resulting in a total of 18 participants.
Each participant was instructed to draw one line drawing for each of the eight tasks.
The participants' ages ranged from 22 to 39, with a median age of 26.
We aimed to recruit people from the general public and artists.
Three participants reported working in visual arts.
We reasoned that artists with experience or even a formal education in visual arts might prefer certain techniques or refrain from others.
However, we could not find substantial differences between the line attributes used by artists and non-artists; thus, we combined the results for further analysis.
Given the small sample size of artists, further research would be needed to compare differences between artists and non-artists.
}

\subsection{Tasks}

\begin{table}
\centering
    \caption{Total usage of the line attributes during the drawing study.}
    \footnotesize
\begin{tabular}{llllllll}\toprule
Dashing & Width & Luminance & Waves & Sketchy & Blur & Other \\ \midrule
113 & 77 & 45 & 20 & 11 & 2 & 17\\
\bottomrule
\end{tabular}
    \label{tab:drawing_total}
\end{table}

\begin{table}[tbp]
\centering
    \caption{Types of dashing used in the drawing study.
    }
    \footnotesize
\begin{tabular}{ll}\toprule
Usage & Dashing type \\
\midrule
54 & Single category\\
17 & Longer dashes more certain \\
16 & Categorical dots and dashes\\
11 & Shorter gaps more certain\\
8 & \review{Shorter gaps and longer dashes more certain}\\
7 & Alt. dots and dashes more certain than dashes\\
2 & Alt. dots and dashes less certain than dashes\\
2 & Shorter gaps and shorter dashes more certain\\
1 & Fewer gaps more certain\\
1 & Perp. dashes encode uncertainty\\
\bottomrule
\end{tabular}
    \label{tab:drawing_dash}
\end{table}

The tasks concerned architecture, biomolecules, blood vessels, graphs, hiking maps, precipitation forecasts, and timelines.
\review{We selected application examples from diverse domains to cover common situations where uncertainty is present. However, exhaustive coverage of all possible situations is infeasible for this study setup.
Consequently, the applicability of our results might vary in different domains.
However, we are confident that our coverage is broad enough to provide generalizability to most domains.
In each task, the data had several levels of uncertainty.
All participants received the same set of instructions.
The descriptions were task-specific, and the uncertainty was explained in an easy-to-understand way.
Common for all tasks was that some type of uncertainty had to be encoded in a line drawing by the participants.
Example submissions, including task descriptions for two of the tasks, can be found in Figure \ref{fig:example-submissions}.
The other tasks and some of the answers provided by the participants are in the supplementary material.
}

\begin{figure}[tbh]
    \centering
    \includegraphics[width=0.75\linewidth]{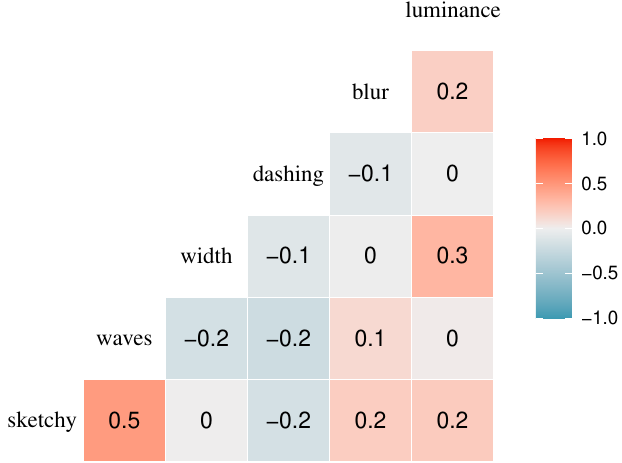}
    \caption{Color-coded Pearson correlation coefficients (phi coefficients) for the drawn line attributes.
    We found that waves were often drawn very sketchy. This is visible in the higher correlation between sketchy lines and waves.
    Width and luminance did also correlate.
    This is probably due to the hand-drawn nature of the lines.
    Higher pressure on the pencil lead to simultaneously wider and darker lines.
    }
    \label{fig:drawing_corr}
\end{figure}

\subsection{Analysis}
In total, we collected 145 line drawings (one participant submitted two alternative drawings for task 2).
\review{
Two of the authors independently reviewed all submissions and classified the line attributes per task and participant into the categories blur, luminance, dashing, sketchiness, waves, width, and other.
}
In many cases, one line drawing was assigned to multiple categories, e.g.,\ width and dashing were used in the same line drawing.
In cases where the classifications differed, the line drawings were discussed and reevaluated.
\review{
To account for the variability in dashing styles, the submissions with dashing were further categorized based on their dashing style.}
Table \ref{tab:drawing_total} shows the total usage of the line attributes during the drawing study.
The four most commonly used line attributes were dashing, width, luminance, and waves.
\review{
Blur was only used twice during the study.
None of the digital submissions included blur. 
Both instances of blurred lines were submitted on paper.
We suspect that blurring might be an intuitive metaphor for uncertainty but is probably harder to draw for most people than, e.g.,\ dashing.
}

We did not find strong trends that favored a line attribute for specific tasks only.
Other techniques, different from the ones we predetermined before evaluating the study, were used 17 times.
Often they were arguably not line attributes, such as the usage of question marks at very uncertain lines.
One participant exclusively drew short lines at feature points for very uncertain lines.
Most of the other techniques were used in tasks with spatial/positional uncertainty, i.e.,\ in the tasks concerning maps and blood vessels.
Similar to ensemble visualization, some participants used multiple lines to indicate uncertainty about the true path of a line.
In another case, a participant indicated possible areas in which the true path could lie by drawing boxes---again, arguably not really a line drawing.
In the task concerning blood vessels, some participants used line drawings that symbolized the area of possible wall locations.
This was one of the tasks where a participant used blur.
Other participants used visualizations similar to error bars (see Figure \ref{fig:error_bars}).
More examples are in the supplementary material.

Table \ref{tab:drawing_dash} shows the dashing techniques used in the drawing study.
By far, the most commonly used dashing technique was dashing as a single category.
In this case, dashing was used alongside other techniques to symbolize one level of uncertainty.
If more than one level was displayed using dashing, most often, the dashes would decrease in length to symbolize higher uncertainty. 
Another common choice was to use dashes and dots as two separate categories.
Some participants combined line attributes.
Very wide lines could, for example, represent very certain data, ``normal'' lines a medium level of certainty, and dashed lines were very uncertain.
We examined if some combinations of line attributes were favored and appeared together extraordinarily often.
Figure \ref{fig:drawing_corr} shows a color-coded correlation matrix for the drawn line attributes.
We find a negative correlation coefficient for wide and dashed lines.
The highest correlation---approx. 0.5---can be found between sketchiness and waves.
During the classification process, we often had to assign sketchiness and waves to the same line, as the waves were often drawn in a sketchy fashion.
This indicates that both line types are fairly similar for hand-drawn lines.
We tried to label a line as a wave if the pattern was relatively regular, resembling a sine wave.
It was labeled sketchy if the pattern was relatively irregular.
However, a clear distinction was often difficult; thus, many lines were labeled as sketchy \emph{and} wavy.
The second-highest correlation (approx. 0.3) can be found between luminance and width.
Again, it was often hard to distinguish between intentionally wide and intentionally dark lines because higher pressure on a pencil leads to simultaneously darker and wider lines.
Both of these higher correlations are probably due to the hand-drawn nature of the lines. 
We could not find any other favored combinations of line attributes.
However, we found that the strongest negative correlations to all other line attributes could be found for dashing.
One reason for this could be that it is probably very easy for the participants to draw dashed lines.
This reduces confounding factors and makes classifications into only this one category more probable.

To summarize, people seem to prefer dashed line attributes for drawing uncertainty.
Other popular choices, in decreasing order, are width, luminance, and waves.
We also found many instances where several line attributes were combined in a single visualization.

\begin{figure}[tb]
    \centering
    \includegraphics[width=0.99\linewidth]{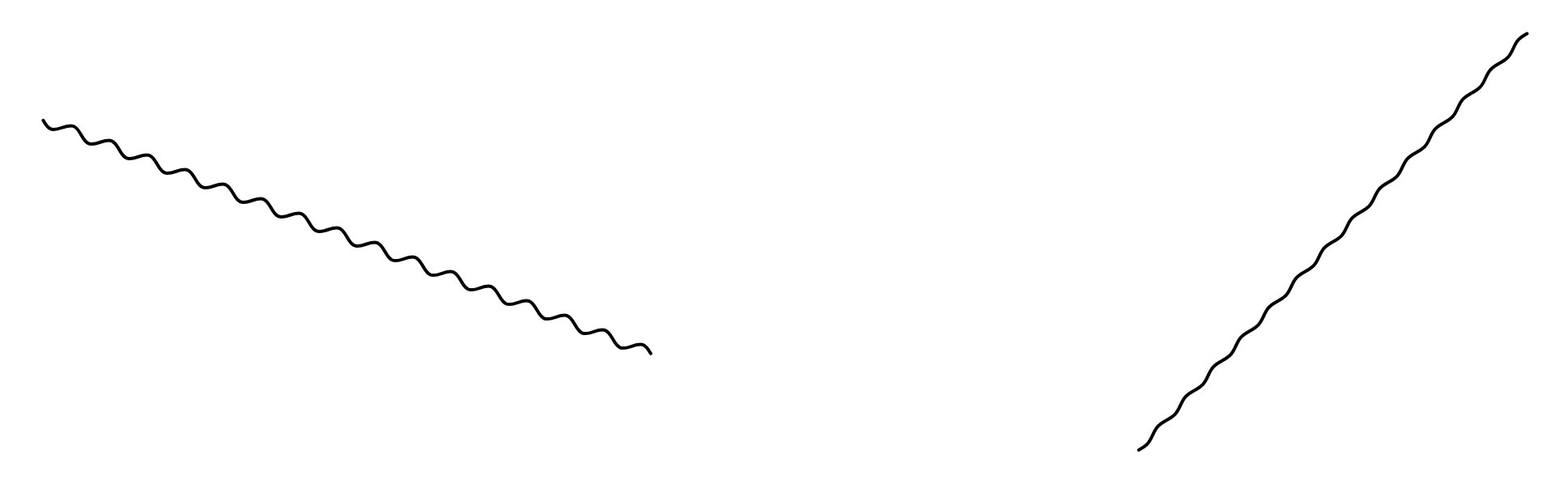}
    \caption{Example task of the JND study.
    \review{
    The participants had to click on the line with the higher amplitude.}
    The images are scaled down.
    }
    \label{fig:jnd_task}
\end{figure}%
\begin{table}[tbh]
\centering
\caption{Standard and test levels of the JND Study.
}
\begin{tabular}[tbh]{llllll}\toprule
Line Attribute & Standard & Tests &&& \\
\midrule
$\textrm{Dashing}_{\textrm{async}}$ & 40 & 36 & 36.7 & 37.3 & 38 \\
\cmidrule{2-6}
{\small length of dash} [px] & 26.7 &  14 & 19.3 & 22 & 24.7\\
&& 28 & 30 & 32 & 36\\
\cmidrule{2-6}
& 13.3 & 5.3 & 8 & 10 & 12 \\
&& 15.3 & 18 & 20.7 & 26 \\
\cmidrule{2-6}
& 0 & 0.7 & 1.3 & 2 & 5.3 \\
\midrule
$\textrm{Dashing}_{\textrm{sync}}$ & 0 & 0.7 & 1.3 & 2 & 2.7 \\
\cmidrule{2-6}
{\small length of dash} [px] & 11.3 & 2.7 & 8 & 9.3 & 10.7 \\
& &  12 & 13.3 & 14.7 & 24 \\
\cmidrule{2-6}
& 24.7 & 12 & 20 & 22 & 23.3 \\
& & 26 & 27.3 & 29.3 & 38 \\
\cmidrule{2-6}
& 38 & 23.3 & 30.7 & 33.3 & 36 \\
\midrule
Luminance [\%]& 0 & 10 & 20 & 30 & 50 \\
\cmidrule{2-6}
& 30 & 0 & 10 & 20 & 25 \\
 & & 35 & 40 & 50 & 70 \\
\cmidrule{2-6}
 & 60 & 30 & 45 & 50 & 55\\
 && 65 & 70 & 75 & 90 \\
\cmidrule{2-6}
& 90 & 70 & 80 & 85 & 88 \\
\midrule
Waves & 0 & 0.3 & 0.7 & 1 & 1.3 \\
\cmidrule{2-6}
{\small amplitude} [px]& 3.3 & 1.3 & 2 & 2.7 & 3 \\
 & & 4 & 4.7 & 6 & 7.3 \\
\cmidrule{2-6}
 & 6.7 & 2.7 & 4.7 & 5.3 & 6  \\
 & & 7.3 & 8 & 9.3 & 11.3 \\
\cmidrule{2-6}
& 13.3 & 6.7 & 9.3 & 10.7 & 12  \\
\midrule
Width [px] & 0.7 & 1.3 & 2 & 2.7 & 3.3 \\
\cmidrule{2-6}
 &6.7 & 4 & 4.7 & 5.3 & 6 \\
& & 7.3 & 8 & 8.7 & 9.3 \\
\cmidrule{2-6}
& 13.3 & 8.7 & 11.3 & 12 & 12.7 \\
& & 14 & 14.7 & 15.3 & 18 \\
\cmidrule{2-6}
& 26.7 & 16.7 & 23.3 & 24.7 & 26\\
\bottomrule
\end{tabular}
\label{tab:jnd-stimuli}
\end{table}

\begin{figure}[tb]
    \captionsetup[subfigure]{justification=centering,singlelinecheck=false}
    \scriptsize
    \begin{subfigure}[b]{\linewidth}
    \centering
    \includegraphics[width=0.9\linewidth]{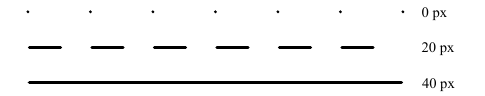}
    \caption{$\textrm{Dashing}_{\textrm{async}}$. 
    \reviewtwo{
    The level annotation describes the dash length $x_1$.
    The gap length $x_2$ can be calculated as $x_2 = 40\,\mathrm{px} - x_1$.
    The dashing style is more closely described in Section \ref{Stimuli_JND}.
    }
    }
    \end{subfigure}
    \vspace{0.1cm}
    
    \begin{subfigure}[b]{\linewidth}
    \centering
    \includegraphics[width=0.9\linewidth]{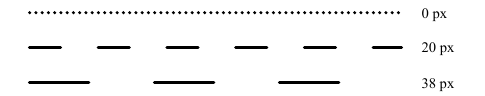}
    \caption{$\textrm{Dashing}_{\textrm{sync}}$. 
    \reviewtwo{
    The level annotation describes the dash length $x_1$.
    The gap length $x_2$ can be calculated as $x_2 = x_1 + 4\,\mathrm{px}$.
    The dashing style is more closely described in Section \ref{Stimuli_JND}.
    }
    % %
    }
    \end{subfigure}
    \vspace{0.1cm}
    
    \centering
    \begin{subfigure}[b]{\linewidth}
    \centering
    \includegraphics[width=0.9\linewidth]{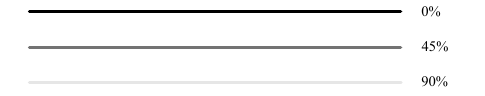}
    \caption{Luminance.}
    \end{subfigure}
    \vspace{0.1cm}
    
    \begin{subfigure}[b]{\linewidth}
    \centering
    \includegraphics[width=0.9\linewidth]{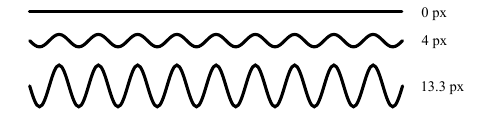}
    \caption{Waves amplitude.}
    \end{subfigure}
    \vspace{0.1cm}
    
    \centering
    \begin{subfigure}[b]{\linewidth}
    \centering
    \includegraphics[width=0.9\linewidth]{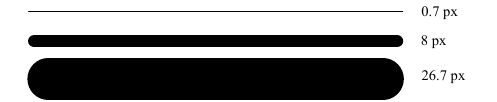}
    \caption{Width.}
    \end{subfigure}
    \caption{Line attributes used in the JND study and for the level recommendations. Displayed are the minima, maxima, and one in-between level for the JND.
    For dashing, 
    dots appear at the minimum level of 0\,px because the rounded cap is not counted as part of the dash length.
    }
    \label{fig:line_samples}
\end{figure}

\begin{figure}[tbp]
    \centering
    \includegraphics[width=0.65\linewidth]{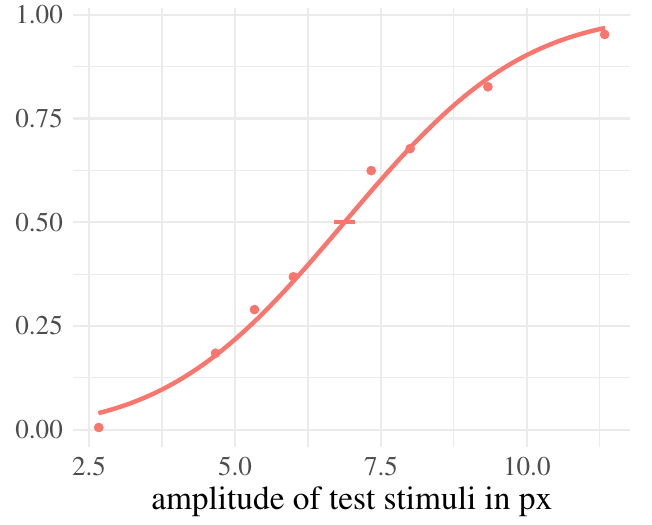}
    \caption{Curve fitting for the 6.7\,px amplitude standard of the wave line type.
    The y-axis shows the proportion of times that participants perceived the test stimulus as bigger than the reference standard.
    The red line represents the fitted GLM.}
    \label{fig:line-fitting}
\end{figure}
\section{Just-noticeable differences}
\label{sec: jnd}
\review{
After our initial exploration of line attributes for uncertainty visualization, we evaluated the number of distinguishable line attribute levels.
For this purpose, we calculated the just-noticeable differences (JND), which in turn, were used to derive level recommendations.
To keep our analysis feasible, we had to limit the number of line variables for which the JND were calculated.
We chose the line attributes most commonly used by the participants of our first study: luminance, waves, width, and dashing.
To account for the high variability of the most popular choice, dashing, we investigated two types of dashing.
For $\textrm{dashing}_\textrm{sync}$ the dashes and gaps grow synchronously, for $\textrm{dashing}_\textrm{async}$ asynchronously.
}
%
%
%----------------------------------------------------------------------------
%
\subsection{Experimental Setup}
The study was set up with a within-participants design.
Per line attribute, four different levels were chosen as reference standards.
For the luminance lines, for example, we chose four different luminance values at which we calculated the JND: 0\%, 30\%, 60\%, and 90\%.
To determine the JND for each of these standards, we used the method of constant stimuli.
This means the participants had to compare different test levels to the standards in two-alternative forced choice tasks.
An example of such a task can be seen in Figure~\ref{fig:jnd_task}.
\review{
In each task, the participants had to click on the line with the higher attribute intensity.
For the wave example in Figure~\ref{fig:jnd_task}, the question would be: ``Which line has a higher amplitude?''.
We included explanations on the meaning of the attributes, e.g.,\ we explained what amplitudes are.
}
\review{
Each pair of lines was shown for 200\,ms.
Typically, visualizations will be visible for a considerably longer time.
However, the overall complexity of a visualization in a real-life scenario is a lot higher than the display of solely two lines.
We chose this time limit to make the tasks more comparable to real-life scenarios, where the comparison of two individual lines should take little time.
}
The participants were under no time constraints for answering, but they were informed that their completion time for each task was being recorded.
The order of the tasks per line attribute was chosen randomly.
\review{
Four test levels were selected above each lowest standard and four test levels below each highest standard.
For the two standards in-between, levels above and below were used.
Consequently, there are four test stimuli for both the minimum and maximum standard and eight test stimuli for the in-between standards.
The JND can be calculated in both directions for the intermediate standards while only in one direction for the maximum and minimum standards.
In each tested direction, two test levels should not be distinguishable (e.g.,\ 10\% and 20\% for the 0\% luminance standard), while the other two should be (e.g.,\ 30\% and 50\%).
}
All standards and tests used during the study can be found in Table~\ref{tab:jnd-stimuli}.
Each participant had to perform each comparison ten times.
We recruited 20 participants.
Thus, there was a total of $\left( 2 \cdot 4 + 2 \cdot 8\right) \cdot 5 \cdot 10 \cdot 20 = 24000$ tasks.
The participation time per participant was approx. 90 minutes, to prevent fatigue and lack of concentration, the participants were regularly reminded to take breaks.
The study was conducted on a 28" LCD-Display with a resolution of $3840 \times 2160\,$px.
Participants individually conducted the study on the same device in the same room with controlled lighting but at different times.
Their distance to the display was approximately 70 cm, and they were instructed to refrain from varying their viewing position substantially.
\review{
We recruited 20 participants, for whom the only criterion for participation was to be at least 18 years old.
The age of the participants ranged from 21 to 31, with a median of 24.5 years.
All participants reported having normal or corrected to normal vision.
}
%-----------------------------------------------------------------------------
%
\subsection{Stimuli} \label{Stimuli_JND}
All stimuli were created on-the-fly as SVG.
While the line attributes were the same for all participants, the line lengths and rotations were generated randomly.
They thus differed for each participant and each repetition.
\review{
This simple randomization approach might introduce confounding factors; however, the extensive sample size and the resulting averaging mitigate potential biases caused by the random generation process.
}
The line lengths ranged from 100 to 800 SVG pixels, and the rotations ranged from 0\degree to 180\degree.
All lines had round line caps and were 2\,px wide (except for the tasks concerning width).
We utilized the full range of our available hardware, which has a device pixel ratio of~1.5. 
Therefore, the smallest unit used in our study is $0.\overline{66}$ logical pixels, which is equivalent to one physical pixel of the hardware used in the study.
All levels are rounded to the first decimal digit.
An overview of the stimuli can be found in Figure~\ref{fig:line_samples}.
The figure shows the minimum, the maximum, and one in-between level for each line attribute.
\review{
The minimum levels were the minimal values possible for the given line attribute.
However, for most line attributes, the maximal values could get infinitely big.
We chose maximal values that we believe to be in a sensible range for most visualizations.
}

\review{
In the following, we describe the line attributes used in the study.
For $\textrm{dashing}_{\textrm{async}}$, the dashes and gaps grow asynchronously.
More specifically, the dashes shrink, and the gaps grow for growing uncertainty while the combined size of one gap and one dash is constant.
The resulting lines are continuous for certain data, have long dashes and small gaps for somewhat certain data, and small dashes with large gaps for uncertain data.
This directly corresponds to the category \emph{shorter gaps and longer dashes more certain}, but it also includes three more commonly used principles: \emph{Dots} (very small dashes), \emph{longer dashes more certain}, and \emph{short gaps more certain}.
}
For a given dash length $x_1$, the gap length $x_2$ can be calculated as $x_2 = 40\,\mathrm{px} - x_1$.
The lengths given correspond to the dash lengths without the rounded caps.
Therefore, even for a dash length of zero, a dot with a diameter of 2\,px is visible.
The $\textrm{dashing}_{\textrm{sync}}$ lines are generated similarly to previously studied lines~\cite{Boukhelifa2012, Sterzik2022}.
Their gaps and dashes grow synchronously for higher uncertainties.
\review{
This is equivalent to the \emph{shorter gaps and shorter dashes more certain} category, which two people used during the drawing study.
}
The gap length can be calculated from the dash length as $x_2 = x_1 + 4\,\mathrm{px}$.
This is because the lengths given correspond to the line lengths without the round cap.
If the cap length was included in the line length, the dash and gap lengths would be the same.
For the line attribute luminance, we vary the luminance in HSL space.
%
% The color ranges from black to light gray (a luminance of 90\%).
%
All wavy lines have equal wavelengths (4\,px) and varying amplitude.
The line attribute width is varied from 0.7\,px to 26.7\,px.

\subsection{Analysis}
\label{sec: JND analysis}
The supplementary material includes all participant responses.
We excluded one participant's responses from the analysis because they seemed to answer randomly.
Their mean accuracy was only 0.596 (95\%~CI~[0.568, 0.624]), much lower than the mean accuracy of all other participants (0.812, 95\%~CI~[0.807, 0.817]).
This excluded participant performed poorly on easy comparisons but better on hard comparisons, reinforcing our belief that their answers were random.

As described by Balestrucci et al.~\cite{Balestrucci2022}, we fitted generalized linear models (GLM) with probit link functions to the data of the individual standards to calculate the JND.
An example of the curve fitting can be found in Figure~\ref{fig:line-fitting}.
In this figure, the dots represent the proportion of times the participants perceived the test as bigger than the standard.
The line represents the fitted GLM.
From the fitted curves, the JND can be calculated as $
    JND = \frac{0.675}{\beta},
    $
where 0.675 is the 75th percentile of the normal distribution and $\beta$ is the slope of the GLM~\cite{Balestrucci2022}.
We repeated this procedure for all standards of every line type.
The resulting JND were rounded to the first decimal place for all line types using pixel units and to the next integer for luminance.
\begin{figure*}[tb]
\centering
    \begin{subfigure}[t]{0.3\linewidth}
    \centering
    \includegraphics[width=\linewidth]{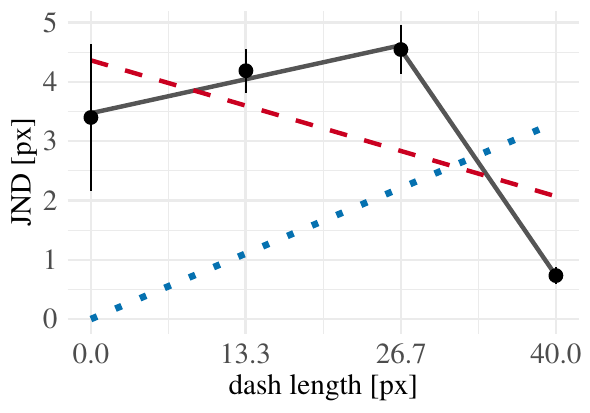}
    \caption{$\textrm{Dashing}_{\textrm{async}}$}
    \label{fig:jnd_dashed1}
    \end{subfigure}
    \hfill
    \begin{subfigure}[t]{0.3\linewidth}
    \centering
    \includegraphics[width=\linewidth]{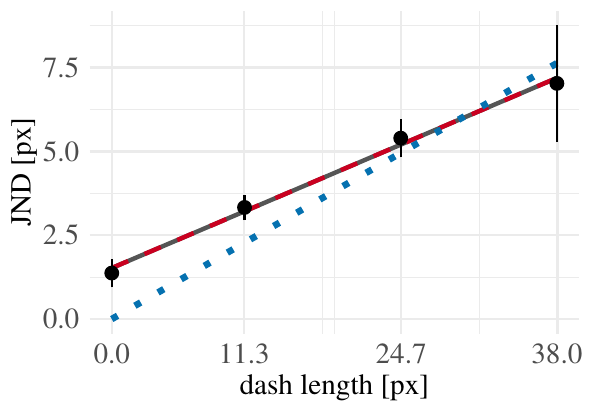}
    
    \caption{$\textrm{Dashing}_{\textrm{sync}}$}
    \label{fig:jnd_dashed2}
    \end{subfigure}
    \hfill
    \begin{subfigure}[t]{0.3\linewidth}
    \centering
    \includegraphics[width=\linewidth]{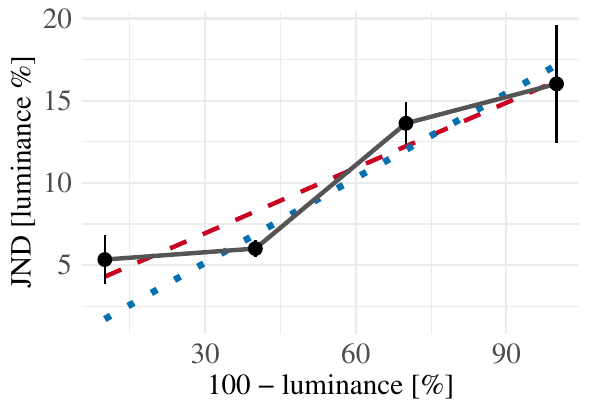}
    \caption{Luminance}
    \label{fig:jnd_value}
    \end{subfigure}
    \begin{subfigure}[t]{0.3\linewidth}
    \centering
    \includegraphics[width=\linewidth]{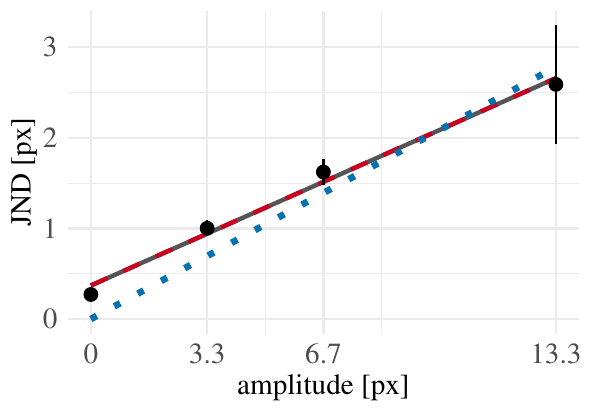}
    \caption{Waves}
    \label{fig:jnd_waves}
    \end{subfigure}
    \hfill
    \begin{subfigure}[t]{0.3\linewidth}
    \centering
    \includegraphics[width=\linewidth]{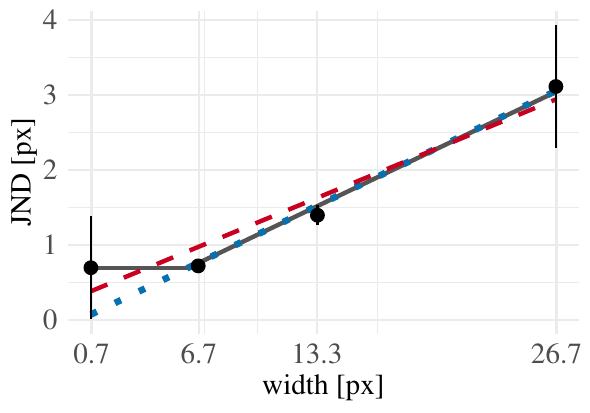}
    \caption{Width}
    \label{fig:jnd_width}
    \end{subfigure}
    \hfill
    \begin{subfigure}[t]{0.3\linewidth}
    \centering
  %     \def\svgwidth{\linewidth}
  % \import{gfx/jnd/}{legend.pdf_tex}
      \scriptsize
    \includegraphics[width = \linewidth]{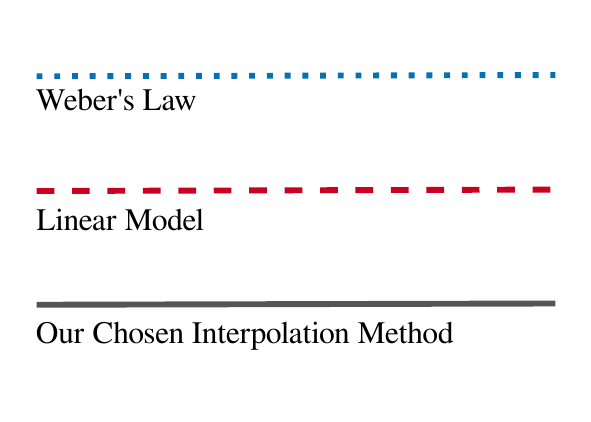}
    \end{subfigure}
    \caption{\reviewtwo{The subfigures show the JND with corresponding confidence intervals in black.
    For deriving level recommendations, it is necessary to interpolate between the individual JND values. Our first choice of interpolation method is Weber's Law (dotted blue lines). Where Weber's Law did not fit the data well, we used linear models (dashed red lines). In some cases, we had to use separate linear models for several ranges of stimulus intensity.
    Our chosen interpolation methods are displayed in dark gray.
    Please refer to Section \ref{sec: JND analysis} for a closer description of our selection process for the interpolation methods per line attribute.}
    }
    \label{fig:jnd_interpolation}
\end{figure*}
Figure~\ref{fig:jnd_interpolation} shows the calculated JND for every line attribute and every standard.
For the width standard of 13.3\,px in Figure~\ref{fig:jnd_width}, the JND is approximately 1.5.
This means that we would need a line with a width of 11.8\,px or smaller, or a line with a width of at least 14.8\,px for people to notice that the width changed compared to the standard of 13.3\,px.
Based on these JND, we derived level recommendations for creating visualizations that encode scalar values using line attributes.
Each level should be one JND apart from the last level.
Therefore, our level recommendations are an upper boundary to the maximal number of levels that people could distinguish for a given line attribute.
Because we only determined the JND at four points in our given range, we need to interpolate between the JND of the different standards for each line attribute.
The interpolation methods are presented in Figure~\ref{fig:jnd_interpolation}.
Our first choice of interpolation method would be Weber's law which states that the ratio between the JND and a stimulus intensity $I$ is constant: $
    \frac{\mathrm{JND}}{I} = K
$
\cite{Munzner2014}.
However, Weber's law is typically not applicable over the whole range of stimulus intensity~\cite{Kay2016}.
Therefore, we often had to resort to other kinds of interpolation, usually linear models.
In contrast to an interpolation following Weber's law, those linear models can have non-zero offsets.
For $\textrm{dashing}_{\textrm{async}}$ (Figure~\ref{fig:jnd_dashed1}), we used a linear model for the first three data points and linear interpolation between the last two data points.
The JNDs for $\textrm{dashing}_{\textrm{async}}$ seem to get larger for the first three standards before decreasing for the last standard.
We hypothesize that the participants rely more on the length of dashes to distinguish individual lines and only start to use the gaps as indicators for the levels when the dashes are considerably longer than the gaps.
Based on our results, we cannot exactly tell at which dash length this change in perception happens, and our linear interpolation between these two points is probably only a very rough approximation of the true values.
In future work, the JND between dash lengths of 26.7\,px, and 40\,px could be examined more closely.
We used a linear model to derive the levels for the $\textrm{dashing}_{\textrm{sync}}$ line attribute (see Figure~\ref{fig:jnd_dashed2}).
The results for luminance (Figure~\ref{fig:jnd_value}) can neither be described accurately using Weber's law nor by fitting a linear model with a non-zero offset.
Other research also shows that a linear model is only a good approximation for high intensities~\cite{Treisman1964}.
Therefore, we chose piecewise linear interpolation between the JND.
In Figure~\ref{fig:jnd_waves}, a linear model describes the wave data well and was, thus, used to determine the levels.
Except for the first standard, the JND for width (Figure~\ref{fig:jnd_width}) seem to follow Weber's law, with a Weber fraction of 0.11, 95\% CI [0.10, 0.13].
Between the first and the second standard, the JND stays approximately constant (0.7$\,$px).
Thus, we determined the levels for the width variable by using constant increments of 0.7$\,$px up to the width of 7$\,$px.
Afterwards, the Weber fraction was used to derive the levels.
Table~\ref{tab:levels} shows the resulting level recommendations for all line attributes.

\begin{table*}[tb]
\centering
\footnotesize
\caption{Perceptually distinct line attribute levels as derived from the results of the JND study.
}
\begin{NiceTabular}{lllllllllllllll}[colortbl-like]\toprule
Line Attribute & \multicolumn{13}{l}{Levels} \\ \midrule
\rowcolor{gray!10}
Luminance [\%] & 0 & 16 & 31 & 44 & 54 & 62 & 68 & 74 & 80 & 86 & 91 &&& \\
$\textrm{Dashing}_{\textrm{async}}$ [px] & 0 & 3.5 & 7.1 & 10.9 & 14.8 & 18.9 & 23.2 & 27.7 & 31.9 & 34.9 & 37.1 & 38.7 & 39.8 &\\
\rowcolor{gray!10}
$\textrm{Dashing}_{\textrm{sync}}$ [px] & 0 & 1.5 & 3.2 & 5.2 & 7.5 & 10.1 & 13.1 & 16.6 & 20.6 & 25.2 & 30.5 & 36.6 & 43.6 &\\
Waves [px]& 0 & 0.4 & 0.8 & 1.3 & 1.9 & 2.6 & 3.4 & 4.4 & 5.5 & 6.8 & 8.3 & 10.1 & 12.2 & 14.7  \\
\rowcolor{gray!10}
\multirow{2}{*}{Width [px]} & 0.7 & 1.4 & 2.1 & 2.8 & 3.5 & 4.2 &  4.9 & 5.6 & 6.3 & 7 & 7.8 & 8.7 & 9.7 & 10.8 \\
\rowcolor{gray!10}
    & 12 & 13.3 & 14.8 & 16.4 & 18.2 & 20.2 & 22.4 & 24.9 & 27.6 \\ 
\bottomrule
\end{NiceTabular}
\label{tab:levels}
\end{table*}
\begin{figure*}[tbh]
    \scriptsize
    \centering
\begin{subfigure}{0.45\linewidth}
    \centering
    \includegraphics[width = \linewidth]{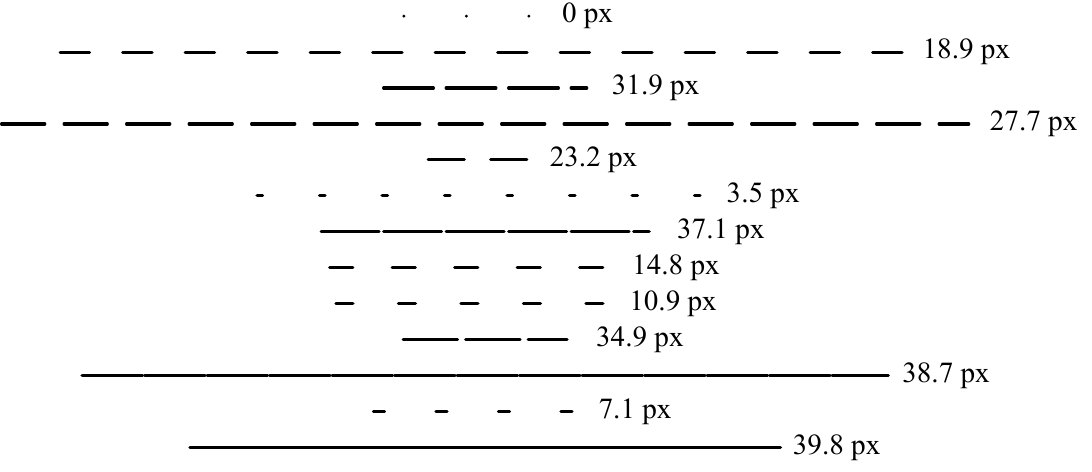}
    \caption{Unsorted}
\end{subfigure}
\begin{subfigure}{0.45\linewidth}
    \centering
    \includegraphics[width = \linewidth]{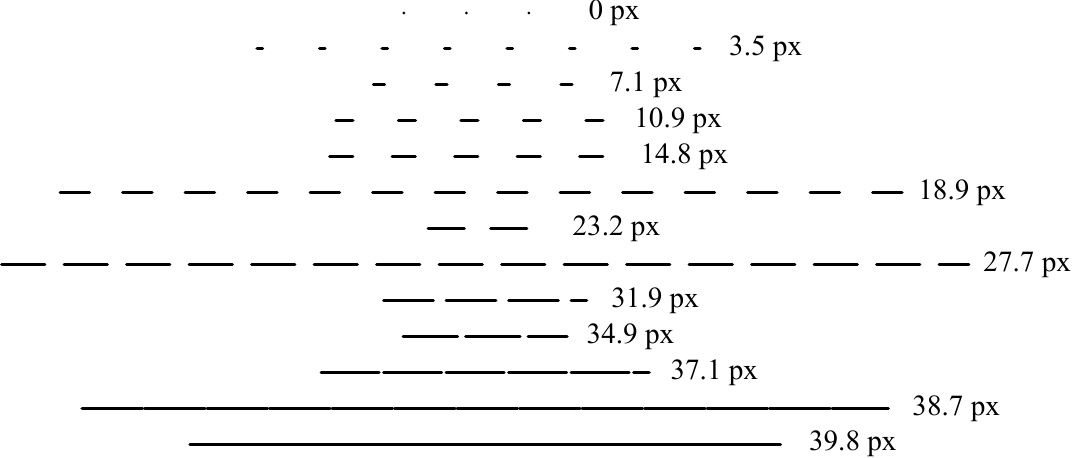}
    \caption{Sorted}
\end{subfigure}
\caption{Ranking task for the $\textrm{Dashing}_{\textrm{async}}$ lines.
The participants saw the unsorted lines (left side) and were instructed to sort them using drag-and-drop (right side).
Here, the size of the gap length is given next to the corresponding lines for reference.
These annotations were not present during the study.
In addition to describing the directionality of the sorting in the task description, the first and last line were fixed, to the smallest and largest level, respectively.
The images have been scaled down.
}
    \label{fig:levels_task}
\end{figure*}

\section{Ranking}
\label{sec: ranking}
In our last perceptual evaluation, we verified the derived levels.
\review{
We implemented an online study consisting of a ranking task for each line attribute evaluated in the JND study.
}
The participants were instructed not to use mobile devices to ensure comparable conditions.
%

%
%----------------------------------------------------------------------------
%
\subsection{Experimental Setup}
The study was conducted online.
In the ranking tasks, the participants had to order the recommended levels of the respective line types per drag-and-drop.
To prevent confusion, we specified the correct order twice:
In the textual description of the task and secondly, the first and last lines were
fixed in place.
\review{
For the width line attribute, for example, they were asked to ``arrange the lines by ascending width''.
Additionally, we specified that the first and the last line are fixed in place and that the thinnest line is on top, the widest at the bottom.
}
An example of one of the tasks is shown in Figure \ref{fig:levels_task}.
The participation time was approximately 10 minutes.
\review{
We recruited 54 participants.
The minimum participation age was again 18 years.
There were no other exclusion criteria.
The considered participants’ ages ranged from 18 to 57 years, with a median age of 23.5 years.
One participant reported a color vision deficiency, but our study did not use colors.
Apart from this, all participants reported having normal or corrected to normal vision. 
Sixteen participants were experienced in the field of visualization.
Their experience ranged from one to 30 years, with a median of 1.5 years.
}
\subsection{Stimuli}
Again, the stimuli were created as SVG on-the-fly. 
The lines had a width of two pixels, except for the line type width.
We randomized the initial order, the length of the lines, and the order of the line types to prevent learning effects.
Besides, we used rounded caps and varied the line length to prevent comparisons based only on the edges. 
The line attribute levels were derived in Section \ref{sec: jnd} and can be found in Table \ref{tab:levels}.
\subsection{Analysis}
We used Spearman's rank correlation coefficient $\rho$ to compare the participants' ratings to the ground truth.
It evaluates the strength of the monotonic relationship between two ordinal variables and can therefore be used to determine the correlation or agreement between two ratings.
$\rho$ ranges from -1 to 1 from strong negative correlations to strong positive correlations~\cite{Cunningham2012}.

If our predicted levels from the JND study are sensible, we expect to find coefficients close to 1 during this study.
We excluded the results of two participants from the study because it is very probable that they either did not understand the task correctly or did not try to solve it.
They answered very fast---under two seconds in some cases---and their rankings were consistently very far from the ground truth and the answers of the other participants.
Their Spearman coefficients were smaller than 0.5 in all but one instance and even negative in one case.
In general, we found very high coefficients.
For each line attribute but width, more than 75\% of the participants ordered every item correctly.
For the width attribute, only 42\% achieved a perfect score.
However, this was to be expected as there were approximately two times as many distinguishable levels predicted as for the other attributes.
Even so, many participants still achieved nearly perfect rankings with an average of $\rho = 0.997$, 95\% CI [0.996, 0.998].
Summary statistics for all line attributes can be found in Table \ref{tab:level_eval}.

\begin{table}[b]
\caption{Spearman's rank correlation coefficient for the level evaluations with 95\% confidence intervals of the mean computed with the bootstrap percentile method.}
\label{tab:level_eval}
\centering
\footnotesize
\begin{tabular}{lllll}\toprule
Line attribute & Minimum & Mean & 95\% CI \\ \midrule
$\textrm{Dashing}_{\textrm{async}}$ & 0.900 & 0.995 & [0.990, 0.999] \\
$\textrm{Dashing}_{\textrm{sync}}$ & 0.991 & 0.998 & [0.998, 0.999]\\
Luminance & 0.833  & 0.993 & [0.984, 0.999] \\
Waves & 0.979 & 0.999 & [0.998, 1.000] \\
Width & 0.983 & 0.997 & [0.996, 0.998] \\
\bottomrule
\end{tabular}
\end{table}

\section{Discriminability Results}
Comparing the results of the individual line attributes is very difficult.
The suitability of a line attribute largely depends on the task at hand.
If there are many long, continuous lines in a visualization, dashing might work very well, as it would be possible to utilize long dash lengths.
However, if the lines are relatively short, it is impossible to faithfully depict dashed lines with long dashes and/or long gaps.
This is especially true for the $\textrm{dashing}_{\textrm{sync}}$ line.
They require a lot of space when the dashes get large because the gaps and dashes grow simultaneously, whereas the combined size of one dash and one gap stays the same regardless of dash length for $\textrm{dashing}_{\textrm{async}}$ lines.
Nonetheless, we did not find any major difference regarding discriminability for both styles of dashing when using dashes up to approximately 40\,px for our lines.
We found 13 perceptually distinct levels for both types of dashing.
The number of usable levels for the luminance attribute seems to be limited to 11 when using white backgrounds, as we did.
If the background has other or varying colors, the number of usable levels is probably even smaller.
When using a wavy line attribute, the resulting lines need more space perpendicular to the line direction. 
Our level recommendations consider lines with a diameter of up to approx. 30\,px.
In this range, we found 14 recommendable levels.
For the line attribute width, we also considered lines with a diameter of up to 30\,px and found 23 distinguishable levels.
While the lines are very thin, very small changes in width can be detected reliably.
This is considerably more than for wavy lines, even though the diameter is the same.
Line types with more perceptually distinct levels can be used for tasks that need a more fine-grained distinction between scalar values.

\begin{figure}[tb]
    \centering
    \includegraphics[scale = 0.24]{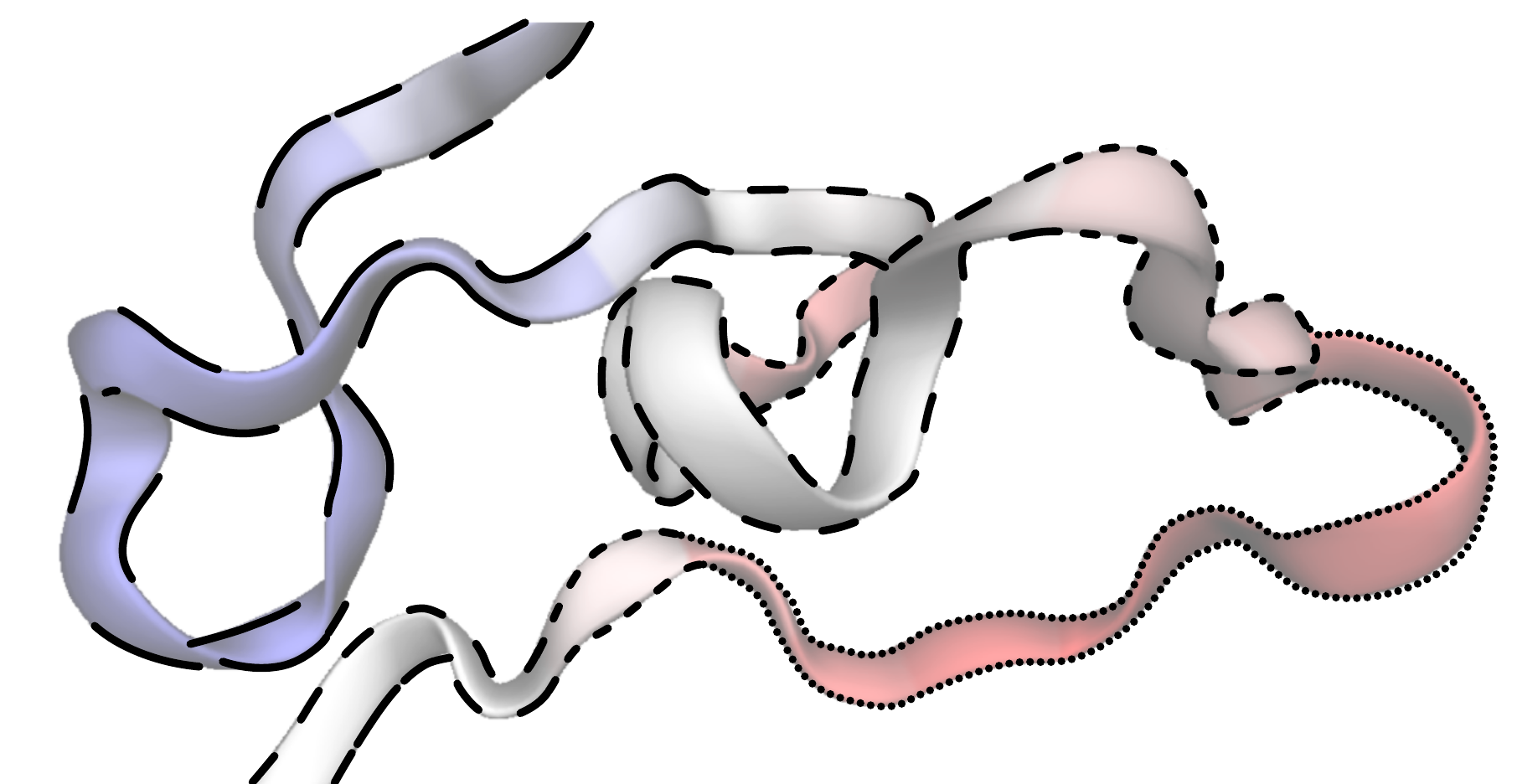}
    \caption{Generating five levels of dashing using our $\textrm{dashing}_{\textrm{sync}}$ level recommendations for the molecular task of study 1.
    Similar to Figure \ref{fig:mol-task}, the colors encode the positional uncertainty of the molecule.
    The line image has been scaled down by a factor of 2.}
    \label{fig:molecule-lines}
\end{figure}

Of course, our concrete level recommendations can only be applied directly in tasks similar to ours, where the line lengths are long enough, there are no other visual distractions, and the lines are close to each other.
Additionally, while the lines were pseudo-randomly rotated and only shown for 200\,ms in the second study, this was not true for the third study, making the task easier.
However, we are confident that our level recommendations are a good starting point for finding suitable levels in specific visualization tasks.
The more the concrete visualization conditions differ from our "perfect" conditions, the fewer levels should be used.
Figure \ref{fig:molecule-lines} shows the molecule from Task 5 of the drawing study.
This time, five levels of dashing were created based on our level recommendations.
We chose every third level to approximate a perceptually uniform spacing.

\section{Discussion and Future Work}
\review{
The results of our drawing study indicate preferences for specific line attributes for drawing uncertainty.
The participants probably chose the line attributes based on several criteria, such as intuitiveness and ease of drawing.
The most prevalent line attributes we found were dashing, luminance, waves, and width.
However, in the literature, there exist other types of line attributes that are used for encoding uncertainty, most notably blur and sketchiness~\cite{Sterzik2022, Sterzik2023, Boukhelifa2012, Guo2015, Gortler2018}.
It would be incorrect to assume that the line attributes we have investigated closely are the only common line attributes.
Our drawing study was specific to hand-drawn lines and uncertainty visualization.
For other types of data and computer-generated lines, other line attributes might be more common.
In future work, additional line attributes found in other domains could be explored further.
Our work could be complemented by a literature-driven approach to find line attributes that are common in visualization research and practice.
Additionally, a more extensive drawing study could be conducted.
The participants could be interviewed about their motives for using specific line attributes for specific scenarios.
To more specifically evaluate line attributes for intuitiveness, the work of Boukhelifa et al.~\cite{Boukhelifa2012} of presenting line drawings with blurred or sketchy lines and asking about interpretations could be extended for additional scenarios and other line attributes.
}
\review{
Regarding our level recommendations, it is important to acknowledge some limitations.
We calculated the Just Noticeable Difference (JND) at four reference standards for each line attribute.
However, the interpolations required to derive the line recommendations may have limited accuracy in certain cases.
In future work, it would be valuable to conduct a more detailed evaluation of specific regions of attribute intensity.
Especially for $\textrm{dashing}_\textrm{async}$, it is essential to closely examine the region around the maximum to accurately identify the true maximum and provide precise line level recommendations.
}
Additionally, the proposed line levels we found based on our study results might only hold under our test conditions.
In scenarios where direct comparisons between the lines are not as easy, the differences between the levels need to be bigger.
However, we think that our recommended levels can be seen as a baseline for ideal conditions; therefore, the calculated JND can be used to derive levels suitable for specific circumstances with less favorable conditions.
We limited ourselves to the five most popular options, but even for the styles we chose, there would be other possible line attributes to vary.
For waves, one could, for example, vary the frequency instead of the amplitude, and---as our first study showed---there is an abundance of other dashing varieties that we could not examine further.
Nevertheless, we are confident that our evaluated line attributes and attribute levels provide a varied foundation for designing easily readable line-based visualizations in many scenarios.

In contrast to prior work by Boukhelifa et al.~\cite{Boukhelifa2012}, we found more perceptually distinct levels.
They reported finding three levels for dashing and four for blur, luminance, and sketchiness.
Our ranges of possible values differ somewhat, but even if we restrict our ranges to theirs, we find at least twice as many values.
As they mentioned in their paper, their selection is somewhat conservative and based on the assumption that each level will be independently compared to the min/max values.
They used numeric estimation methods to compare the perception of a line attribute and its physical value.
Afterwards, they manually selected levels and then determined if these levels are perceptually distinct by doing ANOVA and cluster analysis.
Our study was specifically designed to calculate the JND, leading to our results being more fine-grained than theirs.

\review{
Our findings are in line with evaluations by Sterzik et al.~\cite{Sterzik2022, Sterzik2023}, focusing on line attributes for encoding uncertainty on molecular surfaces.
Specifically, they discovered that encoding uncertainty as line width facilitated an accurate perception of differences in uncertainty.
However, they observed slightly inferior but comparable results when using luminance encoding.
These findings suggest that, despite having fewer distinguishable levels in our study, luminance could still be a viable choice for specific scenarios.
Furthermore, they found that dashing exhibited lower discriminability compared to other investigated line variables.
Their study employed contour lines on molecular models, which presented challenges such as discontinuous and short lines.
This stresses the importance of considering the specific application scenario while choosing a line attribute for encoding information.
}

\review{
Future work should investigate the impact of background colors on line attribute perception.
We suspect that colored backgrounds considerably influence the perception of line attributes, specifically luminance.
The line length could be another important factor for the perception of line attributes---especially width---and dashing.
We attempted to produce sensible results for a wide range of line lengths by randomizing the line length; however, an investigation of the effect of the line length on the perception of the other attributes would be an interesting direction for future work.
Furthermore, combinations of several line attributes need to be examined more closely.
This exploration can address questions regarding the extent to which combining line attributes improves the ability to distinguish between different data values.
Additionally, it can explore the relative intuitiveness of different attribute combinations, the perception of attribute combinations as separate channels, and the identification of preferences for specific combinations for specific tasks.
}

\section{Summary and Conclusion}
We conducted three studies on the perception of line attributes.
\review{
We tried to answer the question \emph{how do people draw lines that encode uncertainty} in our first study.
We found dashing, luminance, waves, and width to be the most prevalent line attributes for drawing uncertainty.
}
\review{
In the second part of our evaluations, we determined the JND for the line attributes $\textrm{dashing}_{\textrm{sync}}$, $\textrm{dashing}_{\textrm{async}}$, luminance, waves and width.
We had to limit the number of investigated line attributes to make our experiments feasible.
The line attributes were chosen based on their prevalence in our first study.
}
Using the calculated JND, we derived level recommendations for all five line attributes. 
We confirmed the chosen levels using drag-and-drop-style ranking tasks in an online study.
We found that most participants were able to clearly distinguish the proposed attribute levels (see Table~\ref{tab:level_eval}).
For width, we found 23 distinguishable levels, which is considerably more than for the other line attributes.
Therefore, width would be a good choice if the visualization has enough space to include wide lines.
Alternatively, if the wide lines occlude too much of the underlying visualization, wavy lines could be used. 
We found 14 distinguishable levels for waves.
Dashing was a very popular choice among our drawing study participants.
During the JND study, we found 13 distinguishable levels for both dashing types that we tested.
It should only be used if the lines present in the visualization are long and continuous.
Otherwise, long dashes cannot be displayed correctly.
We found 11 levels for luminance and, therefore, only recommend it if few levels need to be displayed.
\review{Our study results provide practical guidance for researchers seeking appropriate line attributes to encode information, particularly uncertainty information.
Drawing from the findings of our initial study, researchers can identify commonly used line variables or explore less conventional but more specific encodings, such as perpendicular lines for area representations.
By following our level recommendations, researchers can create visualizations that maintain consistent perceptual differences between levels, enabling effective information communication.
}
\review{To conclude, we are confident that our studies can provide a basis for more informed information and particularly uncertainty encoding in line attributes.
We identified popular line attributes for encoding uncertainty and give recommendations on the attribute levels that can be used to facilitate a correct perception of differences.}

\acknowledgments{%
This work was partially funded by the Deutsche Forschungsgemeinschaft (DFG, German Research Foundation) –- Project-ID 437702916.
}

\bibliographystyle{abbrv-doi-hyperref}

\bibliography{line-study}

%% ^^^^^ EVERYTHING HERE DOES NOT COUNT FOR THE CONTENT PAGE LIMIT (UP TO 2 PAGES) ^^^^^ %%

% \appendix

\end{document}